# *Review Article*
# Emerging classes of antioxidant to cancer therapy: a review of clinical and experimental studies

Qurat-ul-Ain,[1*] and M. Iqbal Choudhary[1,2]

[1]Dr. Panjwani Center for Molecular Medicine and Drug Research, International Center for Chemical and Biological Sciences, University of Karachi, Karachi-75270, Pakistan
[2]H. E. J. Research Institute of Chemistry, International Center for Chemical and Biological Sciences, University of Karachi, Karachi-75270, Pakistan

## ABSTRACT

Oxidative stress is an imbalance between reactive oxygen species(ROS) production and antioxidant defense system and considered to be involved in the etiology of various type of diseases including cancers. Although, since last many decades several antioxidants belonging to the various synthetic and natural classes and plant extracts have been tested in clinical trials against oxidative stress induced various diseases however, these clinical trials end with antioxidants undesirable effects. In this review, we will describe the most recent findings in the oxidative stress field, highlighting sources of free radical's production and its related oxidative damage at cellular and molecular level, we will also have described new and existing classes of antioxidants and free radical scavengers and their related clinical trials.

## Contents

## Key Words

## 1. Free Radicals and Oxidative Damage: Chemistry, Cell and Molecular Biology

Atoms, molecules or ions with outer shell unpaired electron are called free radicals. Unpaired electrons are capable of altering an atom's or a molecule's chemical reactivity, thereby making it more reactive towards other molecules (Lü, Lin, Yao, and Chen, 2010). In biological systems, these free radicals originate from oxygen, nitrogen, and sulphur during a number of biochemical processes, and in normal and abnormal cell metabolism (Sies, 2018). All these free radicals inside the cells form a group of molecules called reactive oxygen/nitrogen/sulfur species (ROS/RNS/RSS). One of the important groups of free radicals is reactive oxygen species (ROS) that induces transformation of superoxide anion ($O_2^{\cdot-}$), hydroperoxyl radical ($HO_2$), hydrogen peroxide ($H_2O_2$), singlet oxygen ($^1O_2$), and hydroxyl radical ($OH^{\bullet-}$) (Carocho and Ferreira, 2013). Mitochondrial electron transport chain (ETC) is the main source of superoxide anion. However, free radicals are also produced by peroxisomes, and during the process catalyzed by enzymes, such as xanthine oxidase, and NADPH oxidase.

However, several cell metabolic and physiological processes, such as phagocytosis, inflammation, and physical exercise also initiate free radical production (Gào and Schöttker, 2017). Mitochondrial complex I, complex II, and complex III that reside at the inner mitochondrial membrane can leak electrons that react with molecular oxygen, and start ROS production by generating superoxide (Loschen *et al.*, 1974; Sabharwal and Schumacker, 2014; Simula, Nazio, and Campello, 2017).

Superoxide stays in the compartment where it is generated due to its low membrane permeability. However, despite the low membrane permeability, small amounts of superoxide, and hydrogen peroxide can diffuse from the mitochondria *via* voltage dependent anion channel (VDAC), and oxidize one or more cysteine residues; hence, altering protein structure and activity (Belhadj Slimen *et al.*, 2014). Various physiological processes are controlled by redox regulation (Bak and Weerapana, 2015). The reactions leading to the production of ROS are shown in Fig. 1.1.

Superoxide anion is generated either through enzymes or metal-catalyzed processes. Superoxide dismutase (SOD) can dismutate the superoxide to hydrogen peroxide, while catalase (CAT) will reduce hydrogen peroxide to water. The Fenton reaction transforms hydrogen peroxide to the hydroxyl radical HO$^{\bullet}$ in the presence of iron (Carocho and Ferreira, 2013). In cellular and biological systems, superoxide-driven Fenton and Haber–Weiss reactions play important roles as they generate highly reactive radicals from the poorly reactive ones. ROS, especially hydroxyl radicals, are capble of attacking all biomolecules, thus producing damage, and contributing to the endothelial dysfunction, which is a starting point for cancers. Iron ($Fe^{2+}$) may be a possible reason for the formation of ROS, but its importance in this process is still questionable (Formanowicz, Radom, Rybarczyk, and Formanowicz, 2018). A reactive nitrogen species (RNS), peroxynitrite ($ONOO^-$), is formed when the second messenger, nitric oxide ($^{\bullet}NO$) free radical, is produced endogenously by nitric oxide synthases, which reacts with superoxide ($O_2^{\bullet-}$), the resulting $ONOO^-$ further reacts with other chemical entities and form dinitrogen trioxide ($N_2O_3$) and nitrogen dioxide ($^{\bullet}NO_2$). Among all the RNS, only ($ONOO^-$) is highly reactive, and potentially cause damage to macromolecules, since it can enter through anion channels inside the cell where it reacts slowly and selectively with the biological macromolecules throughout the cells, including protein thiol groups, amino acid residues, lipids, and DNA bases (Cortese-Krott *et al.*, 2017). Similar to ROS and RNS, RSS originate from thiols and form a disulfide, either disulfide-*S*-monoxide or disulfide-*S*-dioxide as a result of further oxidation (Giles and Jacob, 2002; Olson *et al.,* 2018). In biological system, RSS, ROS, and RNS also act as signaling molecules.

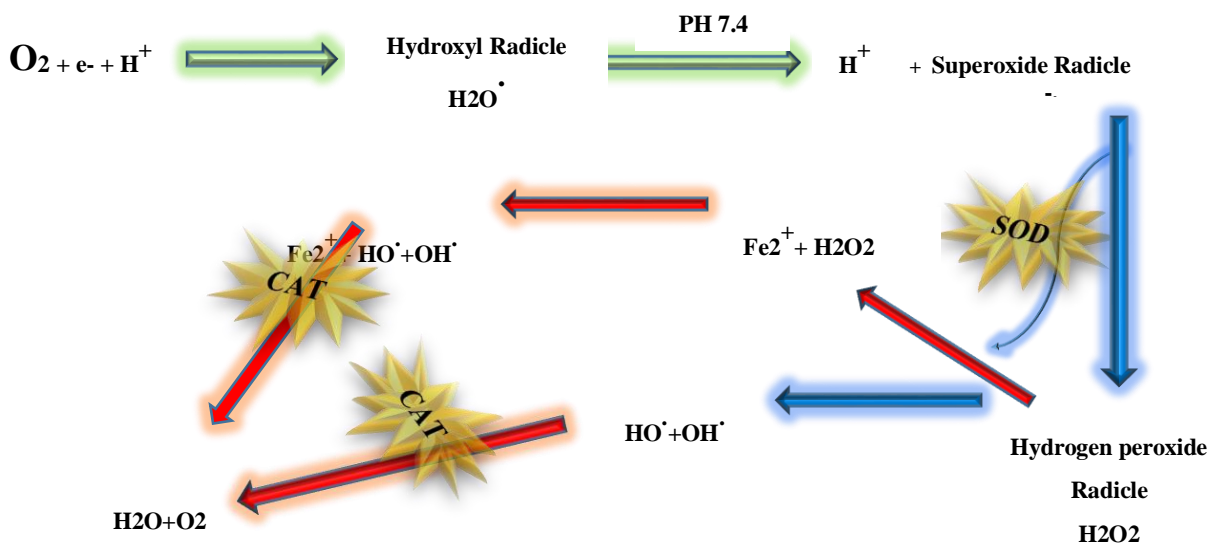

**Fig. 1.1: Overview of ROS production in human cellular systems.** Green arrow shows lipid peroxidation. Haber–Weiss reactions and the Fenton reactions are indicated by blue and red arrows, respectively. Radicals or reactive molecules ($H_2O_2$) are indicated by bold letters. Enzyme catalase and superoxide dismutase are shortened as SOD and CAT, respectively (Carocho and Ferreira, 2013).

## 2. Molecular Targets and Mechanism of Free Radical Damage

Free radicals can react and oxidize all major biological macromolecules in the cells, including deoxyribonucleic acid (DNA), ribonucleic acid (RNA), sugars, lipids, and proteins (Lü, Lin, Yao, and Chen, 2010; Matkarimov and Saparbaev, 2017) (Fig. 1.2). Proteins undergo three distinct types of oxidative modifications: (1) cysteine residues of signaling proteins can be modified to disulfide or to sulfenic acids or to carbonyl derivatives, (2) free radical induced peptide bond cleavage, and (3) cross-linked protein production due to the reaction of protein with lipid peroxidation products (Lobo *et al.*, 2010) (Fig. 1.2). The protein oxidation consequently effects changes in structures and functions of proteins. Hydroxyl radicals can extract a hydrogen atom from methylene carbon of a fatty acid and thereby initiate lipid peroxidation. The lipid radical can undergo molecular rearrangement, and react with oxygen to produce a peroxyl radical. Peroxyl radicals can propagate a chain reaction to form trichloromethyl radical ($CCl_3O_2$) after addition of oxygen to carbon tetrachloride ($CCl_4$) that can attack the lipids (Gaschler and Stockwell, 2017) (Fig. 1.2).

Polyunsaturated fatty acids (PUFAs) are more prone to free radical attack due to their large number of double bonds as compared to monounsaturated (MUFA) and saturated fatty acids (SFA). A minor pathway of lipid peroxidation initiates when singlet oxygen generates lipid peroxides by attacking PUFA's side chains. The end products of lipid peroxidation affect signal transduction as second messengers, and also exert cytotoxic or genotoxic effects (Agmon and Stockwell, 2017; Zielinski and Pratt, 2017) (Fig. 1.2).

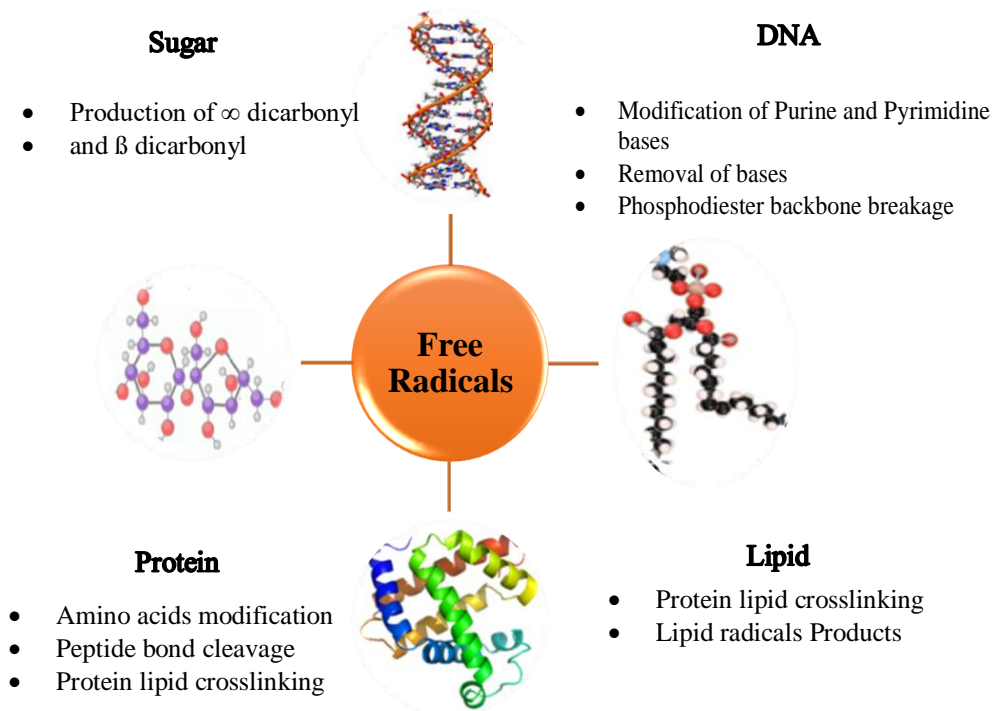

**Fig. 1.2: Molecular targets of free radicals.**

Oxidative damage to DNA includes both chemical and structural changes induced by free radicals. These chemical and structural changes result in characteristic modifications, including strand breaks, base free sites, deletions, base modification, frame shifts, chromosomal rearrangement, and DNA protein cross-links (Fig. 1.2). Hydroxyl radicals can oxidize all components of the DNA molecule, including deoxyribose backbone, and purine and pyrimidine bases. Peroxyl radical also participates in DNA oxidation (Dizdaroglu, Jaruga, Birincioglu, and Rodriguez 2002; Valko, et al., 2004; Chen et al., 2014). ROS induced oxidative damage to DNA triggers a response to recover DNA damage, and limit mutagenesis, cytostasis, and cytotoxicity (McAdam, Brem, and Karran, 2016). Free radicals formed during non-enzymatic glycation result in glycoxidative damage (Fig. 1.2). During glycation, superoxide radicals are produced by autoxidation of short chain molecules, such as glycoaldehyde. Well known mutagens such as ß-dicarbonyls are formed from glycoaldehyde radical by propagation of further chain reaction (Ahmad et al., 2017).

ROS relay cellular signals in a number of cell events by acting as second messengers. However, in bio-molecules, ROS can also induce reversible structural modifications (Li, Jia, and Trush, 2016) that modulate cellular processes. There are several ROS specific signal transduction pathways reported in literature, such as stress-activated/ c-Jun N-terminal protein kinase, NF-KB, AP-1, mitogen and stress-activated kinases (MSK), and p38 group of protein kinases (Marengo et al., 2016).

ROS can also act as physiological mediators in cellular processes in addition to exerting its harmful effects. The redox homeostasis is a balanced state of cellular ROS concentrations (Ursini, Maiorino, and Forman, 2016) achieved by fine tuning of ROS production and ROS detoxifications *via* specific cellular system like NADPH oxidases, and intracellular antioxidant defense system, including thioredoxin, catalase glutathione peroxidase or superoxide dismutases (Hawk and Schafer, 2018).

During different physiological cellular events, such as apoptosis, autophagy, and cell adhesion, the redox status of cells may encounter temporary or stable changes, depending upon the events that take place in a cell. If the redox status of the cell is unbalanced, the cells begin to suffer when ROS production is augmented or/and ROS detoxification is decreased (Sies, 2015).

In addition to a number of diseases, oxidative damage is also believed to contribute to the onset of various types of cancers (Rahman, 2007; Lobo et al., 2010; Rani et al., 2016). However, it is not known if excess oxidation is a source or result of these diseases.

neurological disorders (Autism, Alzheimer's, Parkinson's, and Huntington's diseases), trauma, cardiovascular diseases (atherosclerosis, hypertension, and stroke), renal disorders (glomerulonephritis), liver disorders (Alcoholic liver disease), adult respiratory distress syndrome, auto-immune diseases (lupus erythematous, and rheumatoid arthritis), inflammation, diabetes mellitus, diabetic complications, cataracts, obesity, vasculitis, gastric ulcers, hemochromatosis, and preeclampsia, among others (Rahman, 2007; Lobo, Patil, Phatak, and Chandra, 2010; Rani et al., 2016). As a consequence, new area of research has emerged to understand the antioxidant effects of potential drug candidates.

### 3. Antioxidants and Free Radical Scavengers

Antioxidants are substances of natural or synthetic source that may prevent the process of oxidation of macromolecules by scavenging free radicals. It has also been suggested that they may act by stimulating the expression of free radical quenching enzymes, or by inhibition of free radical producing enzymes (Forman, Davies, and Ursini, 2014). Free radical scavengers may provide new therapeutic strategies for oxidative stress-related diseases. A variety of plant-based free radical scavengers has been described in the literature that are claimed to exert their beneficial effects against free radical induced damage in biological system; some of them are currently used for therapeutic purposes (Dodd et al., 2008). Various fully characterized natural product based antioxidants have been identified; among them are carotenes, estrogens, melatonin, ascorbic acid, tocopherols, uric acid, tocotrienols, glutathione, ubiquinol, bilirubin, lipoic acid, polyphenols, etc. (Singh, Kesharwani, and Keservani, 2018). Apart from the natural ones, numerous non-natural free radical scavengers have been synthesized from previously existing natural templates. They are being tested in cellular and molecular models of oxidative stress for their efficacy (Qurat-ul-Ain, Choudhary, and Scharffetter-Kochanek, 2017).

### *3.1. Pyrimidine Derivatives*

Pyrimidine nucleotide bases are building block of DNA and RNA, and synthetic pyrimidine derivatives exhibit a large number of biological activities (Geng et al., 2018). The pyrimidine analogues decorated with radical scavenging functionalities have been found to counteract oxidative stress induced cancers in many *in vitro* and *in vivo* systems, either by intercalating with DNA nucleotide bases or by reducing free radical induced DNA mutations with subsequent cancer cell growth retardation (Sankarganesh et al., 2017). Many anti-cancer and anti-viral therapies work primarily by these mechanisms (Fig. 1.3) (Ashid et al., 2016; Kaur et al., 2015).

In recent years, anti-cancer drugs are often based on pyrimidine analogues that are either in clinical practice or being evaluated in clinical studies (Cawrse et al., 2018).

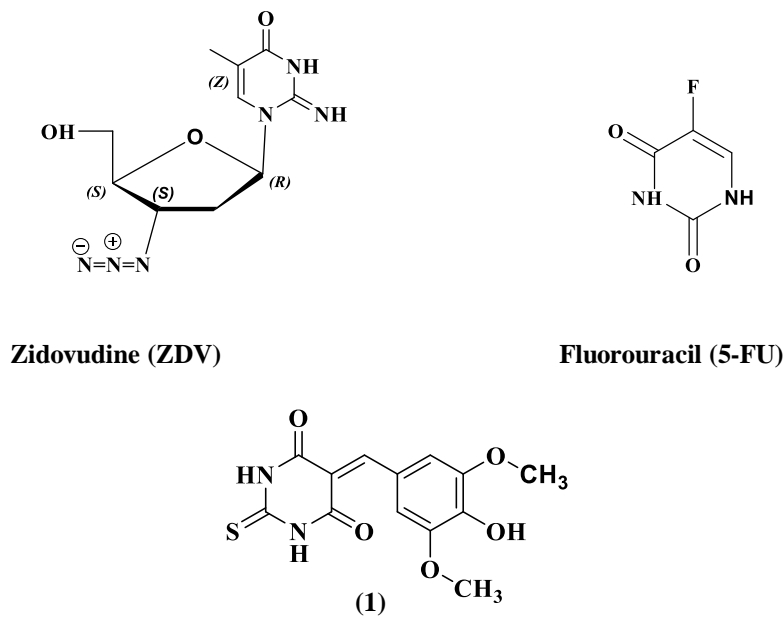

**Fig. 1.3: Structures of pyrimidine based drugs.** Zidovudine is clinically used anti-viral drug, and fluorourical clinically available and orally active anti-cancer agent based on pyrimidine moiety Ashid *et al.*, 2016), structure (**1**) represents newly identified free radical scavenger with anti-proliferation activities (Qurat-ul-Ain *et al.*, 2017).

Identification of antioxidant molecules with DNA binding activities, and investigation of their molecular mechanisms of action towards oxidative damage induced diseases, such as some cancers, can offer new anti-cancer drug candidates (Bano, Kumar, and Dudhe, 2012; Barakat, *et al.*, 2016).

### *3.2. Bis-coumarin Derivatives*

Bis-coumarins are derived from the benzopyrone family of natural compounds.

These compounds are based on a benzene ring joined to a pyrone ring. Their diverse pharmacological properties have attracted immense scientific interest in recent years (Fig. 1.4) (Stefanachi, 2018). Coumarin derivatives are now the subject of intensive research due to their biological activities (Detsi, Kontogiorgis, and Hadjipavlou-Litina, 2017). Recently, in *in vivo* models coumarins have exhibited anti-hepatotoxic activity. They were also found to deplete cytochrome P450 (Singh and Pathak, 2016). Furthermore, a 6-functionalized 1-azacoumarin is an orally active anti-tumor drug, and has recently been tested in human clinical trials in view of its capacity of inhibiting farnesylation in nanomolar concentrations. Due to adverse side effects associated with existing coumarins, there is still a need to synthesize and develop better inhibitors with an improved therapeutic index (Stefanchi *et al*., 2018).

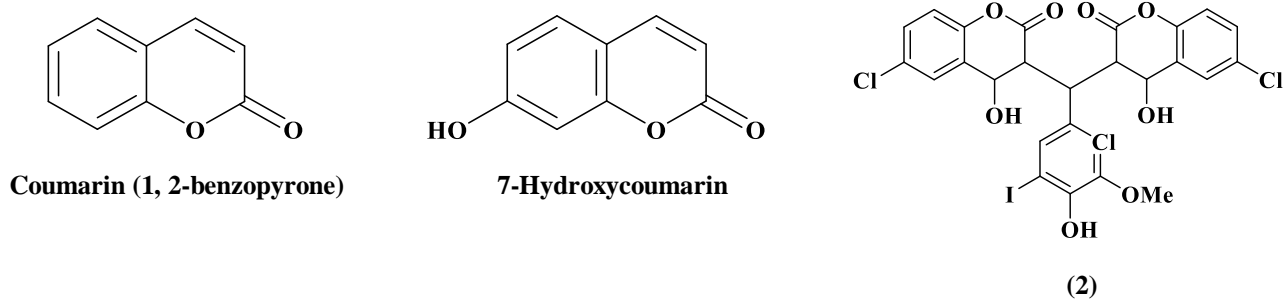

**Fig. 1.4: Structures of coumarin based anti-metastatic/anti-melanoma agents.** Promising anti-metastatic/anti-melanoma agents: Coumarin (1,2-benzopyrone) and 7-Hydroxycoumarin are based on coumarin scaffold, and have been evaluated in phase II studies (Marshall *et al.*, 1994; Mohler *et al.*, 1992), compound (**2**) has been evaluated in oxidative stress cell models (Qurat-ul-Ain *et al.*, 2018).

In various human cancer cell lines, such as HL-60 (leukemia), ACHN (renal), MCF-7 (breast), H727 (lung), and A549 (lung), coumarin and 7-hydroxycoumarin demonstrated growth-inhibitory cytostatic activity. They also have been evaluated in phase II clinical trials against prostate cancer, malignant melanoma, and metastatic renal cell carcinoma (Musa, *et al*., 2008). For example, coumarins with phenolic hydroxyl groups prevent the formation of 5-hydroxyeicosatetraenoic acid (5-HETE) and hydroxyheptadecanoic acid (HHT) in the arachidonic inflammation pathway due to their ability to scavenge reactive oxygen species (Schneider and Bucar 2005).

## 3.3. Thiazole Derivatives

Thiazoles are aromatic compounds with five-membered heterocyclic ring. Due to their antioxidant, anti-convulsant, anti-inflammatory, anti-tumor, anti-hyperlipidemic, anti-hypertensive, insecticidal, and pesticidal activities, thiazole compounds and their synthetic derivatives are used as active ingredients in various pharmaceutical formulations (Khan *et al*., 2016; Salar *et al*., 2016).

A clinically available anti-diabetic agent, Zopolrestat that is built upon thiazole moiety as its active ingredient, is effectively used against diabetic complications. Similarly, a modified thiazole benzene sulfonamides, *i.e.*, *N*-(6-substituted-1, 3-benzothiazol-2-yl) has been found to exhibit its potential as antidiabetic agent in *in vivo* system against NIDDM (non-insulindependent diabetes mellitus) in an animal model (Fig. 1.5) (Moreno-Díaz *et al*., 2008).

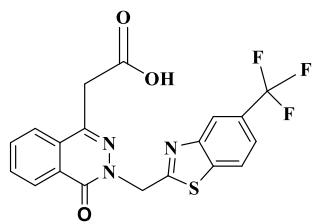
**Zopolrestat**

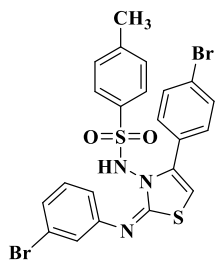
**(3)**

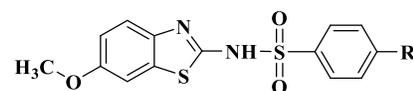
**Benzene sulfonamides**

**Fig. 1.5: Structures of clinically available drugs based on thiazole moiety.** Zopolrestat and benzene sulfonamide *N*-(6-substituted-1,3-benzothiazol-2-yl) have been synthesized based on thiazole moiety. Zopolrestat is a clinically available anti-diabetic agent while benzene sulfonamides *N*-(6-substituted-1,3-benzothiazol-2-yl) has been evaluated in preclinical trials as an anti-diabetic agent against NIDDM (Moreno-Díaz *et al*., 2008), compound (**3**) has been evaluated in oxidative stress cell models (Qurat-ul-Ain *et al*., 2017).

## 3.4. Aryl Schiff's Bases

These synthetic compounds contain characteristic –HC=N– group Aryl Schiff's bases were found to exhibit anti-pyretic, anti-fungal, anti-viral, anti-parasitic, anti-bacterial, anti-tubercular, anti-inflammatory, and plant hormone activities. They also act as HDM2-p53 protein-protein, and dual AKT1/2 inhibitors. They display P2X7, MCH1, and NPY5 antagonism (Parcha and Kaur, 2017). Aryl Schiff's bases of benzyldehyde *N*-mustard and 4'-p-aminophenyl thiazole showed significant activity against lymphoid leukemia in phase III clinical trial (Fig. 1.6) (Modi, Sabnis, and Deliwala, 1970).

Recently, various synthetic and natural product based aryl Schiff's bases were found to have antioxidant/radical scavenging activities with consequent protection from oxidative damage in many models of human diseases, including various types of cancers (Malik *et al.,* 2018). Antioxidants have been added as health promoters to diets, cosmetic products, and pharmaceutical preparations. Some antioxidants also have been tested in various *in vivo* clinical trials with undesirable results (Khurana *et al*., 2018).

with undesirable results (Khurana *et al.*, 2018). For instance, "The Alpha-Tocopherol Beta-Carotene Cancer Prevention Study" (ATBC), conducted on heavy smokers and alcohol drinkers, showed that beta-carotene supplementation resulted in 16% more lung cancers, and 14% more deaths (Albanes *et al.*, 1996). Similarly, the "Carotenoid and Retinol Efficacy Trial" (CARET), showed a 28% increase in lung cancer rates in individuals who were given beta-carotene as suppliment. Many more intervention studies of antioxidant molecules have been carried out either with no results or negative outcomes.

More importantly, some of them have indicated that high doses of an antioxidant such as vitamin E may even raise the overall mortality risks (Sesso *et al.*, 2008; Gerss and Köpcke, 2009).

The pharmacokinetics properties (absorption, distribution, and elimination) of these antioxidants are possible reasons for the failure of these antioxidants in the reported *in vivo* studies. Additionally, *in vitro* antioxidants may lose such a property *in vivo* if the functional group in an antioxidant molecule is shielded by proteins or modified. These events make antioxidants behave differently in a chemical or biological environment.

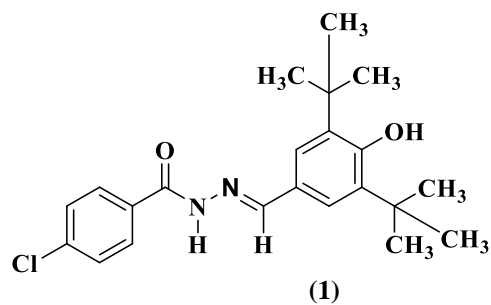
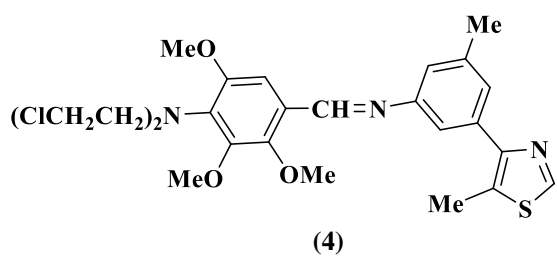

**Fig. 1.6: Structures of potential anticancer aryl Schiff's base.** Anti-leukemic aryl Schiff's bases of benzyldehyde *N*-mustard and 4'-p-aminophenyl thiazole has been evaluated in phase III clinical trial (Modi, Sabnis, and Deliwala, 1970). compound (**4**) has been evaluated in oxidative stress cell models (Qurat-ul-Ain *et al.*, 2017).

Many synthetic and natural radical scavengers/antioxidants can be modified by methylation in the intestinal wall, which results in the loss of their antioxidant properties. As a result, only a fraction of antioxidant molecule, although somewhat bio-available, reaches its target site. For example, RRR alpha-tocopherol that is a well-known natural antioxidant, that due to its ability to bind with tocopherol associated proteins (TAP), is slowly eliminated and thus its presence in the body is significant. Instead, most of the phenolic antioxidants, and for instance, other tocopherols, are taken up in minor amounts, and are extensively eliminated by the drug metabolizing enzymes. The phenolic compounds may lose their antioxidant properties either by binding with the proteins that prevents their interaction with reactive oxygen species (ROS) or by alteration of their antioxidant groups or by rapid metabolism, after absorption (Espinosa and Esposito, 2014). This does not mean that molecules that have lost their antioxidant capacity due to modification of the antioxidant group, or their shielding by proteins in a cellular environment, are biologically inactive because such molecules can exert their function as ligands of receptors or enzymes (Heim, Tagliaferro, and Bobilya, 2013; Qurat-ul-Ain, Choudhary, and Scharffetter-Kochanek, 2018).

Moreover, natural or synthetic antioxidants may possess several non-antioxidant functions that may add to or substitute their native antioxidant properties. In conclusion, both a naturally occurring free radical scavenger or a compound that has been characterized as antioxidant due to having free radical scavenging functionalities, in some situation may not exert an antioxidant function in some situations. The well-known antioxidant alpha-tocopherol, and the mild antioxidants retinol or estrogens may be listed as examples. All these natural antioxidants are protected from oxidation or enzymatic modifications by associating with certain proteins such as the cellular "tocopherol associated protein" (TAP), which prevent their attack by oxygen radicals (Azzi, 2009; Azzi, Meydani, Meydani, and Zingg, 2016). Therefore, free radical molecules that have shown their activity in a chemical environment may have their functions either modified or blocked in biological systems like human body. However, these antioxidant molecules, and their metabolites with masked antioxidant activities might display more important activities than free radical scavenging activitiy in physiological system (Choudhary and Kochanek, 2017). It is, therefore, important to explore other useful biological targets of antioxidants/free radical scavengers along with their radical scavenging properties.

38

## 4. Conclusion

Free radicals produced from different sources are generally known to induced oxidative stress and detrimental to human health. A large number of studies demonstrates that in fact free radicals target all biological macromolecules and contribute to initiation and progression of several pathologies, including various types of cancers.

A number of antioxidants class have been identified that able to counteract oxidative stress in *in vitro* system via their free radical scavenging activity and considered to be mitigate its effects on individuals' health, gained enormous attention from the biomedical research, however some of them when tested in clinical trials showed their negative results due to their metabolism and masking of their free radical functional groups therefore, it is important to explore new classes of antioxidants compounds and their other useful biological targets of antioxidants/free radical scavengers along with their radical scavenging properties that can be well useful to human health, particularly regarding cancer treatment.

## Conflicts of Interest

The authors state no conflict of interest.

## Authors' Contributions

Qurat-ul-Ain: manuscript writing, M. Iqbal review of manuscript.

## Acknowledgments

This work was supported by the International Union of Biochemistry and Molecular Biology (IUBMB)## References

Agmon, E., and Stockwell, B. R. (**2017**). Lipid homeostasis and regulated cell death. *Current Opinion in Chemical Biology*, 39, 83-89.

Ahmad, S., Akhter, F., Shahab, U., Rafi, Z., Khan, M. S., Nabi, R., and Ashraf, J. M. (**2017**). Do all roads lead to the Rome? The glycation perspective! *Seminars in Cancer Biology*.

Albanes, D., Heinonen, O. P., Taylor, P. R., Virtamo, J., Edwards, B. K., Rautalahti, M., and Barrett, M. J. (**1996**). α-Tocopherol and β-carotene supplements and lung cancer incidence in the alpha-tocopherol, beta-carotene cancer prevention study: Effects of base-line characteristics and study compliance. JNCI: *Journal of the National Cancer Institute*, 88 (21), 1560-1570.

Ashid, M., Yogi, P., Katariya, D., Agarwal, P., and Joshi, A. (**2016**). Pyrimidine: Medicinal and biological significance. A review, *World Journal of Pharmacy and Pharmaceutical Sciences*, 5(9) 990-1009.

Azzi, A. (**2009**). How can a chemically well established antioxidant work differently when in the body?. *IUBMB Life*, 61(12), 1159-1160.

Azzi, A., Meydani, S. N., Meydani, M., and Zingg, J. M. (**2016**). The rise, the fall and the renaissance of vitamin E. *Archives of Biochemistry and Biophysics*, 595, 100-108.

Bak, D. W., and Weerapana, E. (**2015**). Cysteine-mediated redox signalling in the mitochondria. *Molecular BioSystems*, 11(3), 678-697.

Bano, T., Kumar, N., and Dudhe, R. (**2012**). Free radical scavenging properties of pyrimidine derivatives. *Organic and Medicinal Chemistry Letters*, 2(1), 34.

Barakat, A., Ghabbour, H. A., Al-Majid, A. M., Imad, R., Javaid, K., Shaikh, N. N., and Wadood, A. (**2016**). Synthesis, X-ray crystal structures, biological evaluation, and molecular docking studies of a series of barbiturate derivatives. *Journal of Chemistry*, 2016.

Belhadj Slimen, I., Najar, T., Ghram, A., Dabbebi, H., Ben Mrad, M., and Abdrabbah, M. (**2014**). Reactive oxygen species, heat stress and oxidative-induced mitochondrial damage. A review. *International Journal of Hyperthermia*, 30(7), 513-523.

Carocho, M., and Ferreira, I. C. (**2013**). A review on antioxidants, prooxidants and related controversy: natural and synthetic compounds, screening and analysis methodologies and future perspectives. *Food and Chemical Toxicology, 51*, 15-25.

Cawrse, B. M., Lapidus, R. S., Cooper, B., Choi, E. Y., and Seley-Radtke, K. L. (**2018**). Anticancer properties of halogenated pyrrolo [3,2-d] pyrimidines with decreased toxicity *via N5* substitution. *ChemMed Chem. 13(*2), 178-185.

Choudhary, M. I., and Kochanek, K. S. (**2017**). Modulation of melanoma cell proliferation and spreading by novel small molecular weight antioxidants. *Free Radical Biology and Medicine*, 108, S28.

Cortese-Krott, M. M., Koning, A., Kuhnle, G. G., Nagy, P., Bianco, C. L., Pasch, A., and Olson, K. R. (**2017**). The reactive species interactome: evolutionary emergence, biological significance, and opportunities for redox metabolomics and personalized medicine. *Antioxidants and Redox Signaling*, 27(10), 684-712.

Detsi, A., Kontogiorgis, C., and Hadjipavlou-Litina, D. (**2017**). Coumarin derivatives: An updated patent review (2015-2016). *Expert Opinion on Therapeutic Patents*, 27(11), 1201-1226.

Dizdaroglu, M., Jaruga, P., Birincioglu, M., and Rodriguez, H. (**2002**). Free radical-induced damage to DNA: Mechanisms and measurement 1, 2. *Free Radical Biology and Medicine, 32*(11), 1102-1115.